\documentclass[superscriptaddress,twocolumn,showpacs,preprintnumbers,amsmath,amsmb]{revtex4}

\usepackage{graphicx}
\usepackage{dcolumn}
\usepackage{bm}

\begin{document}

\preprint{APS/123-QED}

\title{Detection of Atmospheric Muon Neutrinos \\with the IceCube 9-String
Detector}

\affiliation{III Physikalisches Institut, RWTH Aachen University, D-52056 Aachen, Germany}
\affiliation{Dept.~of Physics and Astronomy, University of Alaska Anchorage, 3211 Providence Dr., Anchorage, AK 99508, USA}
\affiliation{CTSPS, Clark-Atlanta University, Atlanta, GA 30314, USA}
\affiliation{Dept.~of Physics, Southern University, Baton Rouge, LA 70813, USA}
\affiliation{Dept.~of Physics, University of California, Berkeley, CA 94720, USA}
\affiliation{Institut f\"ur Physik, Humboldt-Universit\"at zu Berlin, D-12489 Berlin, Germany}
\affiliation{Lawrence Berkeley National Laboratory, Berkeley, CA 94720, USA}
\affiliation{Universit\'e Libre de Bruxelles, Science Faculty CP230, B-1050 Brussels, Belgium}
\affiliation{Vrije Universiteit Brussel, Dienst ELEM, B-1050 Brussels, Belgium}
\affiliation{Dept.~of Physics, Chiba University, Chiba 263-8522 Japan}
\affiliation{Dept.~of Physics and Astronomy, University of Canterbury, Private Bag 4800, Christchurch, New Zealand}
\affiliation{Dept.~of Physics, University of Maryland, College Park, MD 20742, USA}
\affiliation{Dept.~of Physics, Universit\"at Dortmund, D-44221 Dortmund, Germany}
\affiliation{Dept.~of Subatomic and Radiation Physics, University of Gent, B-9000 Gent, Belgium}
\affiliation{Max-Planck-Institut f\"ur Kernphysik, D-69177 Heidelberg, Germany}
\affiliation{Dept.~of Physics and Astronomy, University of California, Irvine, CA 92697, USA}
\affiliation{Dept.~of Physics and Astronomy, University of Kansas, Lawrence, KS 66045, USA}
\affiliation{Blackett Laboratory, Imperial College, London SW7 2BW, UK}
\affiliation{Dept.~of Astronomy, University of Wisconsin, Madison, WI 53706, USA}
\affiliation{Dept.~of Physics, University of Wisconsin, Madison, WI 53706, USA}
\affiliation{Institute of Physics, University of Mainz, Staudinger Weg 7, D-55099 Mainz, Germany}
\affiliation{University of Mons-Hainaut, 7000 Mons, Belgium}
\affiliation{Bartol Research Institute and Department of Physics and Astronomy, University of Delaware, Newark, DE 19716, USA}
\affiliation{Dept.~of Physics, University of Oxford, 1 Keble Road, Oxford OX1 3NP, UK}
\affiliation{Institute for Advanced Study, Princeton, NJ 08540, USA}
\affiliation{Dept.~of Physics, University of Wisconsin, River Falls, WI 54022, USA}
\affiliation{Dept.~of Physics, Stockholm University, SE-10691 Stockholm, Sweden}
\affiliation{Dept.~of Astronomy and Astrophysics, Pennsylvania State University, University Park, PA 16802, USA}
\affiliation{Dept.~of Physics, Pennsylvania State University, University Park, PA 16802, USA}
\affiliation{Division of High Energy Physics, Uppsala University, S-75121 Uppsala, Sweden}
\affiliation{Dept.~of Physics and Astronomy, Utrecht University/SRON, NL-3584 CC Utrecht, The Netherlands}
\affiliation{Dept.~of Physics, University of Wuppertal, D-42119 Wuppertal, Germany}
\affiliation{DESY, D-15735 Zeuthen, Germany}

\author{A.~Achterberg}
\affiliation{Dept.~of Physics and Astronomy, Utrecht University/SRON, NL-3584 CC Utrecht, The Netherlands}
\author{M.~Ackermann}
\affiliation{DESY, D-15735 Zeuthen, Germany}
\author{J.~Adams}
\affiliation{Dept.~of Physics and Astronomy, University of Canterbury, Private Bag 4800, Christchurch, New Zealand}
\author{J.~Ahrens}
\affiliation{Institute of Physics, University of Mainz, Staudinger Weg 7, D-55099 Mainz, Germany}
\author{K.~Andeen}
\affiliation{Dept.~of Physics, University of Wisconsin, Madison, WI 53706, USA}
\author{J.~Auffenberg}
\affiliation{Dept.~of Physics, University of Wuppertal, D-42119 Wuppertal, Germany}
\author{J.~N.~Bahcall}
\thanks{Deceased}
\affiliation{Institute for Advanced Study, Princeton, NJ 08540, USA}
\author{X.~Bai}
\affiliation{Bartol Research Institute and Department of Physics and Astronomy, University of Delaware, Newark, DE 19716, USA}
\author{B.~Baret}
\affiliation{Vrije Universiteit Brussel, Dienst ELEM, B-1050 Brussels, Belgium}
\author{S.~W.~Barwick}
\affiliation{Dept.~of Physics and Astronomy, University of California, Irvine, CA 92697, USA}
% ...

\date{\today}

\collaboration{IceCube Collaboration}
\noaffiliation

\begin{abstract}
The IceCube neutrino detector is a cubic kilometer TeV to PeV neutrino
detector under construction at the geographic South Pole.  
The dominant population of 
neutrinos detected in IceCube is due
to meson decay in cosmic-ray air showers.  
These atmospheric neutrinos are relatively 
well-understood and serve as a calibration
and verification tool for the new detector. In 2006, 
the detector was approximately 10\% completed, and we report on 
data acquired from
the detector in this configuration.
We observe 
an atmospheric neutrino signal consistent with expectations, demonstrating
that the IceCube detector is capable of identifying neutrino events.
In the first 137.4 days of livetime,
234 neutrino candidates were selected with an expectation of 
$211 \pm 76.1(syst.) \pm 14.5(stat.)$ 
events from atmospheric neutrinos.  
\end{abstract}

\maketitle

\section{\label{sec:level1}The IceCube Detector}

The IceCube neutrino detector is 
being deployed in the deep ice below 
the geographic South Pole
\cite{icecubeMuonSensitivity}.  When completed,
the detector will consist of two components.  The InIce detector
is a cubic kilometer of instrumented ice
between 1.5 and 2.5 
kilometers below the surface.  
A cubic-kilometer has been long noted
as the required scale to detect astrophysical neutrino sources 
above the atmospheric neutrino background
(see, e.g. \cite{gaisserHalzenStanev}\cite{waxmanGRBNeutrinos}\cite{agnstuff}).
The IceTop detector
is a square-kilometer air-shower array
at the surface.  This analysis concerns data from the InIce detector
exclusively.

The InIce detector consists of an array of light-sensitive
Digital Optical Modules
(DOMs) 
\cite{firstYearPerfomancePaper}, deployed
17 meters apart in strings of 60.
Strings are arranged
on a hexagonal grid with a spacing of 125 meters.  
The DOMs house a 10-inch Photomultiplier Tube (PMT)
and electronics to acquire, digitize and time stamp pulse waveforms from
the PMT.  
With a waveform fit for fine timing,
the timing resolution for individual photon
arrivals is expected to be less than 2 nanoseconds 
\cite{firstYearPerfomancePaper}.
In 2006, the DOMs were operated in Local Coincidence (LC) 
with their neighbors, 
meaning that a triggered DOM's waveform was only transmitted to the surface
if an adjacent DOM on the string also triggered within $\pm 1000$ ns.  
The surface data acquisition system forms triggered detector-wide events
if 8 or more DOMs are read out in 5 $\mu s$.  

The detector is being deployed in stages during austral summers from 2004
to 2011. 
After the 2005-2006 season, the InIce detector consisted of 9 strings,
termed IC-9.


IceCube is optimized for the detection of muon neutrinos in the TeV to PeV
energy range.  It
is sensitive to these muon neutrinos (and muon anti-neutrinos) by detecting
Cherenkov light from the secondary muon produced when the neutrino
interacts in or near the instrumented volume. 
Neutrino-induced muons are separated from air-shower-induced muons 
by looking
only for muons moving upward through the detector. Up-going muon events
must be the product of a neutrino interaction near the detector, 
since neutrinos are the
only known particles that can traverse the Earth without interacting.

\section{\label{sec:level1}Atmospheric Neutrinos}

Neutrinos produced in cosmic-ray air showers at the Earth are known as 
atmospheric neutrinos and form the chief background to 
potential astrophysical neutrino observation. 
The
atmospheric neutrino spectrum is relatively 
well-understood 
\cite{lotsOnAtmosphericNeutrinos}\cite{atmosphericNeutrinoUncertainties} 
and has
been measured up to $10^{5}$ GeV by 
AMANDA \cite{amandaAtmosphericNeutrinos}.  
Atmospheric neutrinos from the decay
of charmed mesons 
can contribute 
significantly above $10^{4}$ GeV, depending on the model
(see e.g. \cite{costaPromptComponent}\cite{bugaevPromptReference}\cite{Martin:2003us}).  
This prompt component
is not well known due to uncertainties in the charmed meson production, but
with the present exposure of IC-9, this prompt component is negligible and
it is presently neglected.

\section{\label{sec:level1}Results}

Data acquired from the IC-9 detector in 2006 between June and November has
been searched for up-going neutrino candidates.  
The search proceeds by a series of cut levels intended to remove
down-going events as shown in Table \ref{tab:eventPassingRates}.  Initially,
hit cleaning is applied which removes all DOM hits which fall out of a 4 
$\mu s$ time window, and all DOM hits without another DOM hit within a radius
of 100 meters and within a time of 500 ns.  After hit cleaning, we re-trigger,
insisting that at least 8 DOM hits survive hit cleaning.
Simple first-guess reconstruction algorithms
running at the South Pole were used to filter out clearly down-going events.  
Events with fewer than 11 DOMs hit were also filtered to meet bandwidth
constraints from the South Pole.  
The remaining events were transmitted to the data center in the Northern
hemisphere
via satellite and constitute the filter level of the analysis.
At the data center, we reconstructed the direction of events 
using a maximum-likelihood technique similar to the AMANDA muon 
reconstruction
\cite{muonReconstructionNIMPaper}.  
Events which
were reconstructed as down-going
were discarded.
Despite the fact that remaining events appear up-going, 
the data is still dominated 
by misreconstructed down-going events.
These down-going 
events are removed by additional quality cuts.  Events
which pass these quality cuts 
constitute the neutrino candidate dataset. 

Simulated events fall into the three categories shown in Table 
\ref{tab:eventPassingRates}.
``Single shower'' events arise from single 
cosmic-ray air showers in the atmosphere above IceCube and result in a
single muon or bundle of collinear muons in IC-9.  ``Double
shower'' events come from two uncorrelated air showers which
happen to occur within the 5 $\mu s$ event window.  
The CORSIKA\cite{corsikaReference} 
air shower simulation program was
used for the simulation of down-going single and double air-shower events.  
Finally, ``atmospheric
neutrino'' events are muon neutrino 
events from pion and kaon decay in air showers in the
northern hemisphere.  
The atmospheric neutrino model 
of \cite{bartolAtmosphericNeutrino} 
and its extension up to TeV energies \cite{teresasCommuncationWithBarr} 
as well as the cross-section 
parametrization of \cite{cteq5Reference} were used to model the up-going muon
rates due to atmospheric neutrinos.  


\begin{table*}
\caption{\label{tab:eventPassingRates}
Event Passing Rates (Hz).  Shown are the
event passing rates through different processing levels for the
simulated event categories and for experimental data.  The trigger level 
comprises the
events triggering the detector after hit cleaning and re-triggering.  
The filter level comprises
events which passed the online filtering conditions.  
Rates are also shown for events which reconstruct as up-going with and without
the final quality cuts applied (see the text for cut definition).
Note that the rates from air-shower
events have been multiplied by $0.90$ so that the simulation and 
data agree at trigger level.  This is consistent with
an approximately 20\% uncertainty in the absolute cosmic-ray flux.
For the final sample, statistical errors are given for the data and
systematic errors are given for the atmospheric neutrino simulation.
}
\begin{ruledtabular}
\begin{tabular}{ccccc}
Criterion Satisfied&Data&Single Shower&Double Shower&Atmospheric Neutrinos\\
\hline
Trigger Level&124.5&124.5&1.5&$6.6\rm{x}10^{-4}$\\
Filter Level&6.56&4.96&0.45&$3.7\rm{x}10^{-4}$\\
Up-going ($S_{cut}=0$)&0.80&0.49&0.21&$3.3\rm{x}10^{-4}$\\
Up-going ($S_{cut}=10$)&$1.97\cdot 10^{-5}\pm 0.12\cdot 10^{-5}$&-&-&$1.77\cdot 10^{-5}\pm 0.63\cdot 10^{-5}$\\
Up-going ($S_{cut}=10$ and $\theta > 120$)&$1.19\cdot 10^{-5}\pm 0.10\cdot 10^{-5}$&-&-&$1.42\cdot 10^{-5}\pm 0.51\cdot 10^{-5}$\\
\end{tabular}
\end{ruledtabular}
\end{table*}

\begin{figure}
\includegraphics{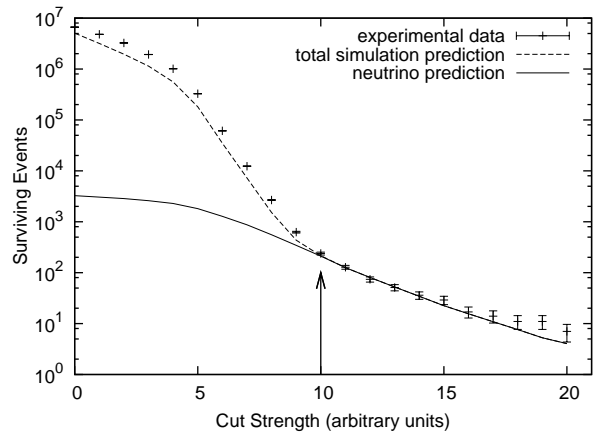}
\caption{\label{fig:dataVsCutStrength}
Data vs Cut Strength.  Shown is the remaining number of events 
as the cut strength $S_{cut}$ (defined in the text) 
is varied.  Curves are shown for the data
and the total simulation prediction.  Also shown is the prediction due to
atmospheric neutrinos alone.  The selection from the text corresponds
to a cuts strength of $S_{cut} = 10$, and is denoted by an arrow.  At
this point, the
data are dominated by atmospheric neutrinos.
}
\end{figure}

The events which are reconstructed as up-going are completely dominated
by down-going muons from single and double-shower cosmic-ray events.
Misreconstructed events 
are typically 
of low quality as measured by two parameters, the number of
direct hits
$N_{dir}$, and the direct length $L_{dir}$.  A direct hit is a photon arrival
in a DOM which is detected between -15 and +75 ns of the 
time expected from the reconstructed muon with no scattering.  
$N_{dir}$ is the total number of direct
hits in an event.  The direct length $L_{dir}$ represents the length of
the reconstructed muon track along which direct hits are observed.  
An event with
a large number of direct hits and a large direct length is a better quality
event because the long lever arm of many unscattered photon arrivals 
increases confidence in the event reconstruction.
The strength of
the quality cuts can be
represented by a dimensionless number $S_{cut}$ which corresponds to 
cuts of $N_{dir} \geq S_{cut}$ and $L_{dir} > 25 \cdot S_{cut}$ meters.  
In addition to these quality cuts, we impose a cut requiring that events
have no more than 46 DOMs hit, which eliminates only about 1\% of the final
event sample.  The purpose of this cut is to leave the high-multiplicity
data blinded for an anticipated search for a high-energy diffuse 
neutrino flux.
Figure \ref{fig:dataVsCutStrength} shows how many events remain
as we turn the cut strength up and increase the signal-to-noise ratio.
The accurate simulation of mis-reconstructed down-going events requires
excellent modeling of both the depth-dependent ice properties and 
DOM sensitivity.  In this
initial study we observe a 60\%-80\% rate discrepancy for misreconstructed
events up to a cut level of about $S_{cut}=8$ or so.  Nevertheless, over
more than four orders of magnitude, the background simulation tracks the
data, the number of wrongly reconstructed tracks is reduced, 
and for $S_{cut}\geq 10$, the data behaves as expected for atmospheric 
neutrinos.
From simulation, we expect neutrinos with energies 
between about $10^2$ and $10^4$ GeV, 
peaked at 1000 GeV, to survive the analysis cuts.

In 137.4 days of livetime we expect $211 \pm 76.1(syst.) \pm 14.5(stat.)$ 
atmospheric neutrino 
events to survive at $S_{cut} = 10$ and 234 events are measured.  
Above a zenith of 120 degrees, where the background contamination is 
small, we measure 142 events with an expectation
of $169 \pm 60.9(syst.) \pm 13.0(stat)$ due to atmospheric neutrinos.
The principal systematic uncertainty in this
atmospheric neutrino expectation is due to the approximately $30\%$ 
theoretical uncertainty
in the atmospheric flux normalization \cite{atmosphericNeutrinoUncertainties}.
The other significant systematic error
is due to uncertainties introduced by the modeling of 
light propagation and the detection efficiency of IceCube DOMs.
The uncertainty in the atmospheric neutrino rate due to this
modeling
is estimated at 20\% and is obtained in this initial study 
by examining changes in the neutrino passing rate when varying the 
cuts to account for the background simulation disagreement 
in Fig. \ref{fig:dataVsCutStrength}.

\begin{figure}
\resizebox{3.5in}{!}{
\includegraphics{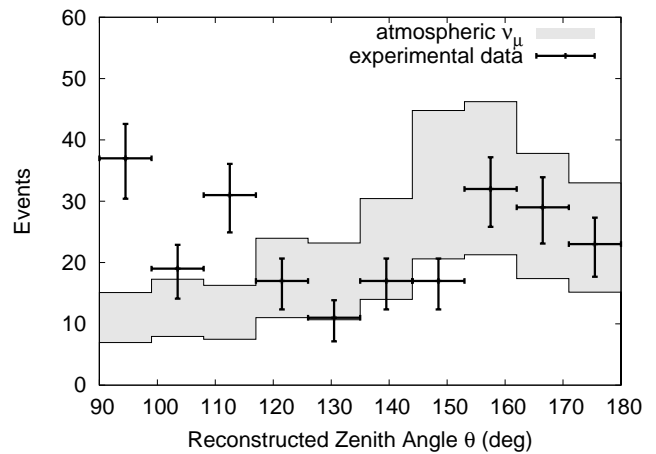}
}
\caption{\label{fig:zenithDistribution}
Distribution of the reconstructed zenith angle $\theta$
of the final event sample.  
A zenith of 90 degrees indicates a horizontal event, and a zenith of 180 
degrees
is a directly up-going event.
The band shown for the atmospheric neutrino simulation 
includes the systematic errors from the text, and the error bars on the
experimental data are statistical.  
Note that uncertainty due to the atmospheric neutrino
flux is an uncertainty in normalization and is nearly 
independent of zenith angle.
}
\end{figure}

\begin{figure}
\resizebox{3.5in}{!}{
\includegraphics{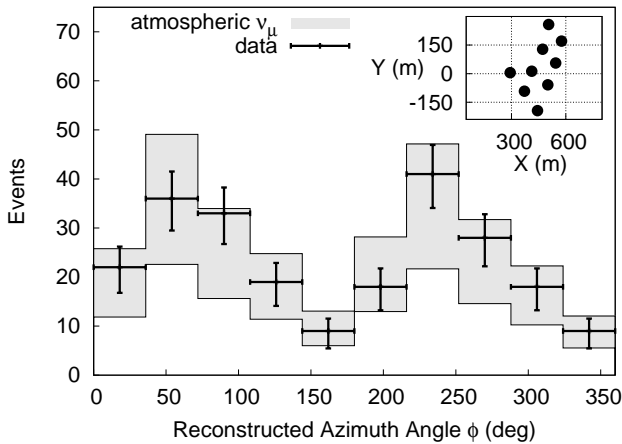}
}
\caption{\label{fig:azimuthDistribution}
Distribution of the reconstructed azimuth angle $\phi$ 
the final event sample.
The band shown for the atmospheric neutrino simulation 
includes the systematic errors from the text, and the error bars on the
experimental data are statistical.  The inset shows the horizontal 
locations of the strings making up 
IC-9 relative to the center of the future array.
Note that uncertainty due to the atmospheric neutrino
flux is an uncertainty in normalization and is nearly 
independent of zenith angle.
}
\end{figure}

Figure \ref{fig:zenithDistribution} shows the measured 
zenith distribution for the final event sample along with the atmospheric
neutrino prediction.  
The zenith angle distribution agrees well with atmospheric neutrino simulation
for vertical events above about 120 degrees.  
The observed excess is
believed to be residual contamination from down-going single
and double cosmic-ray muons.  
This excess disappears if we tighten the cuts beyond $S_{cut}=10$,
suggesting that the recorded events at the horizon are of typically lower 
quality than expected from atmospheric neutrino simulation.  This 
reinforces the belief we are seeing residual background at the horizon.
Above about $S_{cut} = 12$, with low statistics, the data at the horizon are
consistent with a pure atmospheric neutrino signal.
Figure \ref{fig:azimuthDistribution} shows
the azimuth distribution with the IC-9 geometry in the inset.
The azimuth distribution has two strong peaks 
corresponding to the long horizontal axis of the IC-9 detector.  
The cut of 250 meters
on event
length constrains near-horizontal 
events that can be accepted along the short axis of 
IC-9 since the string spacing is 125 meters.  We expect
more uniform azimuthal acceptance in future seasons as the detector
grows and becomes more symmetric.

\begin{figure}
\includegraphics{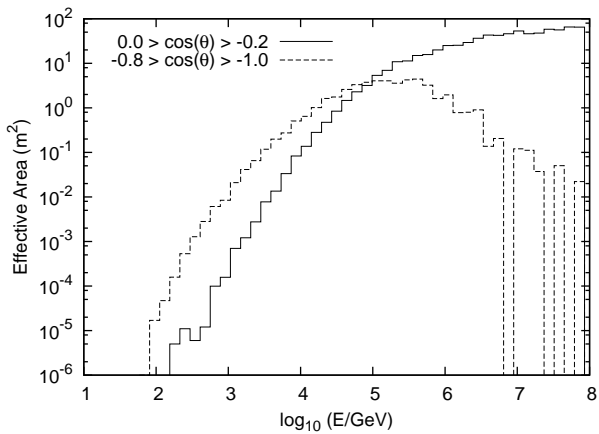}
\caption{\label{fig:effectiveArea}
The effective area of IC-9 to a diffuse source of muon neutrinos.  The 
event selection from the text was used, except that there was no 
multiplicity requirement imposed on the simulated events.
The effective area to muon neutrinos
and muon antineutrinos have been averaged to produce the figure.
The effective area is shown as a function
of neutrino energy, both for vertical and horizontal neutrino events.
Horizontal events have a zenith of 90 degrees and vertical up-going events
have a zenith of 180 degrees.
}
\end{figure}

We can characterize the response of the detector to neutrinos with
an effective area $A_{\it eff}$ which is a function of neutrino energy $E$
and neutrino zenith angle $\theta$.  
The function $A_{\it eff}(E,\theta)$ is defined as the function which
satisfies 
\begin{equation}
R = \int dE \int d\Omega \cdot \Phi(E,\theta) \cdot A_{\it eff}(E,\theta)
\end{equation}
where $\Phi(E,\theta)$ is an arbitrary diffuse neutrino flux and
$R$ is the corresponding rate of 
events surviving analysis cuts.  Figure \ref{fig:effectiveArea} shows
the effective area of IC-9 to neutrinos 
with the event selection of $S_{cut} = 10$, both for neutrinos near the horizon
and for nearly vertical neutrinos.  The effective area to neutrinos
is much smaller than the geometrical area of the detector, due to the
smallness of the neutrino cross-section.
Above $10^{5}$ GeV, the Earth starts to become 
opaque to neutrinos, and the highest energy up-going neutrinos can only be
detected at the horizon.

\section{\label{sec:level1}Conclusions}

In 2006, IceCube was approximately 
10\% deployed and acquiring physics-quality data.  
Atmospheric neutrinos serve as an irreducible background to astrophysical
neutrino observation, as a guaranteed source of neutrinos for calibration
and verification of the detector, and may be studied as a probe of hadronic
interactions at energies inaccessible to terrestrial labs.
In the first 137.4 days of livetime we have identified
234 neutrino candidates with the IC-9 detector.
For events above 120 degrees, this neutrino sample is consistent with
with expectations for a pure atmospheric neutrino sample.
Selection of events was done within six months
of the beginning of data acquisition, demonstrating the viability of the full 
data acquisition chain, from PMT waveform capture at the DOM with 
nanosecond timing, to event selection at the South Pole
and transmission of that selected data via satellite to the North.  
During the 2006-2007 season, 13 more strings were deployed, bringing
the total number of strings for the InIce detector to 22.
The deployment of IceCube will continue during austral 
summers until 2010-2011, while the integrated exposure of IceCube will 
reach a $\rm{km^3} \cdot \rm{year}$ sometime in 2009.

\begin{acknowledgments}

We acknowledge the support from the following agencies:
National Science Foundation-Office of Polar Program,
National Science Foundation-Physics Division,
University of Wisconsin Alumni Research Foundation,
Department of Energy, and National Energy Research Scientific Computing Center
(supported by the Office of Energy Research of the Department of Energy),
the NSF-supported TeraGrid system at the San Diego Supercomputer Center (SDSC),
and the National Center for Supercomputing Applications (NCSA);
Swedish Research Council,
Swedish Polar Research Secretariat,
and Knut and Alice Wallenberg Foundation, Sweden;
German Ministry for Education and Research,
Deutsche Forschungsgemeinschaft (DFG), Germany;
Fund for Scientific Research (FNRS-FWO),
Flanders Institute to encourage scientific and technological research in industry (IWT),
Belgian Federal Office for Scientific, Technical and Cultural affairs (OSTC);
the Netherlands Organisation for Scientific Research (NWO);
M.~Ribordy acknowledges the support of the SNF (Switzerland);
A.~Kappes and J.~D.~Zornoza acknowledge support by the EU Marie Curie OIF Program.

\end{acknowledgments}

\bibliography{bibliography}

\end{document}